\documentclass[reprint, amsmath,amssymb, aps]{revtex4-2}

\usepackage{epsfig, color, ulem}
\usepackage{graphicx}
\def\nabl{\mbox{\boldmath$\nabla$}}
\def\taub{\mbox{\boldmath$\tau$}}
\def\D{\mbox{\boldmath${\cal D}$}}
\def\G{\mbox{\boldmath${\cal G}$}}

\begin{document}

\preprint{APS/123-QED}

\title{Three-dimensional vector waves in time-variant materials, with applications in time-reversed focusing and Green's function retrieval}

\author{Kees Wapenaar$^{1,}$}\thanks{Contact author: c.p.a.wapenaar@tudelft.nl}
\author{Dirk-Jan van Manen$^2$}
\affiliation{$^1$Department of Geoscience and Engineering, Delft University of Technology, 2600 GA Delft, The Netherlands\\
$^2$ETH Z\"urich, Institute of Geophysics, Sonneggstrasse 5, 8092 Z\"urich, Switzerland}

\date{\today}

\begin{abstract}
The behavior of wave propagation and scattering in time-variant materials has been studied by many authors, for electromagnetic as well as for mechanical waves.
Most studies consider two-dimensional waves, governed by scalar wave equations. Here we introduce a unified wave equation for three-dimensional
electromagnetic and elastodynamic vector-wave propagation and scattering in homogeneous, time-variant materials.
Using the symmetry properties of this equation, we establish the conservation of net field-momentum density, and  a reciprocity relation between causal and acausal Green's functions.
The latter relation is exploited in the derivation of the counterparts  of classical time-reversed focusing and Green's function retrieval for time-variant materials.
\end{abstract}

                             
\maketitle
Research towards wave propagation and scattering in time-variant materials, which started in the late 1950's \cite{Morgenthaler58IRE},  gathered significant momentum with the introduction
of manufactured dynamic metamaterials \cite{Caloz2020IEEE1, Apffel2022PRL, Moussa2023NP, Engheta2023Science, Kim2024PRL}.
The wave equation for time-variant materials is similar to that for space-variant materials, with the roles of time and space interchanged \cite{Mendonca2002PS, Torrent2018PRB}.
However, because the causality principle is not interchanged for time and space,
and because the number of space dimensions is in general different from the number of time dimensions, 
waves in homogeneous, time-variant materials behave differently from those in inhomogeneous, time-invariant materials.

Although most of the literature on waves in time-variant materials deals with electromagnetic waves, applications for mechanical waves have been presented 
\cite{Kim2024PRL, Torrent2018PRB, Delory2024PRL}. Fink and coworkers \cite{Fink2017EPJ, Bacot2016NP} demonstrate how a time boundary 
scatters diverging water waves backward in space, 
which then converge and focus at the original source position. This is the counterpart of classical time-reversed acoustics \cite{Fink97PhysicsToday}, 
where time-reversed waves, emitted from a space boundary, focus at the source position.

Most studies on waves in time-variant materials consider transverse-electric, transverse-magnetic, transverse-shear, or acoustic waves in two-dimensional (2D) materials,
which are all governed by scalar wave equations. Studies on three-dimensional (3D) vector waves in time-variant materials are sparse and restricted to electromagnetic waves
\cite{Budko2009PRA, Hoop2014WM, Koutserimpas2020IEEE, Stumpf2026SR, Wapenaar2026PIER}.
Here we present a unified formalism for 3D electromagnetic and elastodynamic vector-wave propagation and scattering 
in homogeneous, time-variant materials, and discuss its applications in time-reversed focusing and Green's function retrieval by spatial cross-correlation.

The unifying equation for waves in time-variant materials reads
\begin{eqnarray}
\partial_t{\bf q}={\bf A}{\bf q}+{\bf d},\label{eq1}
\end{eqnarray}
see the 
Appendix for a detailed derivation.
Here $\partial_t$ stands for $\frac{\partial}{\partial t}$, ${\bf q}({\bf x},t)$ is a $6\times 1$ wave field vector as a function of position ${\bf x}=(x,y,z)$ and time $t$,
${\bf d}({\bf x},t)$ is a $6\times 1$ source vector vector, and
 ${\bf A}(t)$ a $6\times 6$ operator matrix, with
\begin{eqnarray}
{\bf q}=\begin{pmatrix}{\bf U}\\{\bf V}_{\rm d}\end{pmatrix},\quad
{\bf d}=\begin{pmatrix}{\bf a}\\{\bf b}_{\rm d}\end{pmatrix},\quad
{\bf A}=\begin{pmatrix}{\bf O} & -\frac{1}{\beta(t)}{\bf I} \\
-\frac{1}{\alpha(t)}\D& {\bf O}\end{pmatrix}.
\label{eq2}
\end{eqnarray}
Here ${\bf I}$ is an identity matrix and ${\bf O}$ a null matrix. All other quantities in these vectors and matrix are defined in Table \ref{table1}, for 3D electromagnetic 
and 3D elastodynamic waves. For elastodynamic waves it is assumed that the ratio $\lambda(t)/\mu(t)$ 
or, equivalently, the ratio $c_P(t)/c_S(t)$, is time-invariant
\footnote{This is a reasonable assumption for the situation in which the time-variance of the parameters is the result of a time-variant stress, applied to a material with microcracks.
For example, experiments reported by Verdon et al. \cite{Verdon2008GEO} show that $c_P$ and $c_S$  exhibit similar trends as a function of applied stress, 
meaning that their ratio varies much less with  stress.}.
Note that operator $\D$ for electromagnetic waves follows from $\D$ for elastodynamic waves by taking the limit $c_P\to 0$.

\begin{table}[h]
\caption{Specification of quantities in equation (\ref{eq2}). 
For electromagnetic waves, ${\bf D}({\bf x},t)$ and ${\bf B}({\bf x},t)$ are the electric and magnetic flux densities, respectively;
 ${\bf J}^{\rm e}({\bf x},t)$ and ${\bf J}^{\rm m}({\bf x},t)$ are the external electric and magnetic current densities, respectively;
and $\varepsilon(t)$ and $\mu(t)$ are the time-variant permittivity and permeability, respectively. 
For elastodynamic waves, ${\bf m}({\bf x},t)$, ${\bf e}({\bf x},t)$ and $\Theta({\bf x},t)$ are the mechanical momentum density, the strain tensor
and the cubic dilatation, respectively (with $\Theta={\rm tr}({\bf e})$);
${\bf f}({\bf x},t)$, ${\bf h}({\bf x},t)$ and $q({\bf x},t)$ are the external force, deformation-rate and volume injection-rate densities, respectively (with $q={\rm tr}({\bf h})$);
$\rho(t)$, $\lambda(t)$ and $\mu(t)$ are the time-variant mass density and Lam\'e parameters, respectively (with $\lambda(t)/\mu(t)$ assumed time-invariant);
and $c_P(t)$ and $c_S(t)$ are the time-variant $P$- and $S$-wave velocities, respectively.}\label{table1}
\begin{tabular}{l|cc}
\hline
& Electromagnetic & Elastodynamic\\
\hline
${\bf U}$ & ${\bf D}$ & ${\bf m}$\\
${\bf V}_{\rm d}$ &$-\nabl\times{\bf B}$ &$-2\nabl\cdot{\bf e}-\frac{\lambda}{\mu}\nabl\Theta$\\
${\bf a}$ &$-{\bf J}^{\rm e}$ &${\bf f}$\\
${\bf b}_{\rm d}$ &$\nabl\times{\bf J}^{\rm m}$ &$2\nabl\cdot{\bf h}+\frac{\lambda}{\mu}\nabl q$\\
$\alpha$ &$\varepsilon(t)$ &$\rho(t)$\\
$\beta$ &$\mu(t)$ &$1/\mu(t)$\\
$\D$ &$-\nabl\times\nabl\times$& $\frac{c_P^2}{c_S^2}\nabl\nabl^t-\nabl\times\nabl\times$\\
\hline
\end{tabular}
\end{table}

The operator  $\nabl\times$ in Table \ref{table1} is a matrix, defined as
\begin{eqnarray}
\nabl\times=\begin{pmatrix}
0 & -\partial_z & \partial_y\\
\partial_z & 0 & -\partial_x\\
-\partial_y & \partial_x & 0
\end{pmatrix}.\label{eq11}
\end{eqnarray}
With this definition it follows that operator $\D$ is a symmetric matrix, i.e., $\D=\D^t$, where superscript $t$ denotes transposition.
This implies for matrix ${\bf A}(t)$, defined in equation (\ref{eq2}),
\begin{eqnarray}
{\bf N}{\bf A}=-{\bf A}^t{\bf N},\quad\mbox{with}\quad {\bf N}=\begin{pmatrix}{\bf O} & {\bf I}\\-{\bf I} & {\bf O}\end{pmatrix}.\label{eqAsymt}
\end{eqnarray}
This symmetry property of matrix ${\bf A}(t)$ makes that equation (\ref{eq1}) is not only useful for modeling the time-evolution of ${\bf q}({\bf x},t)$ through a homogeneous, time-variant material,
but also that its solutions obey specific symmetry properties, essential for time-reversed focusing and Green's function retrieval, as we will see below.

We introduce a $6\times 6$ propagator matrix ${\bf W}({\bf x},t,t_0)$ (also known as the transfer matrix  \cite{Torrent2018PRB, Pacheco2020NP}),
which obeys the  wave equation (\ref{eq1}) without the source term, hence
\begin{eqnarray}
\partial_t{\bf W}={\bf A}{\bf W},\label{eq29x}
\end{eqnarray}
with initial condition ${\bf W}({\bf x},t_0,t_0)={\bf I}\delta({\bf x})$.
Using this matrix, we can express ${\bf q}({\bf x},t)$ for arbitrary $t$, larger (or smaller) than $t_0$, in terms of the initial (or final) condition ${\bf q}({\bf x},t_0)$, as
\begin{eqnarray}
{\bf q}({\bf x},t)={\bf W}({\bf x},t,t_0)*{\bf q}({\bf x},t_0),\label{eq132x}
\end{eqnarray}
where $*$ denotes a 3D spatial convolution. Here it is assumed that there are no sources for ${\bf q}({\bf x},t)$ between $t_0$ and $t$.

In previous work \cite{Wapenaar2026PIER}, we have shown that the propagator matrix for 3D electromagnetic waves in time-variant materials can be expressed in terms of Green's matrices, as
\begin{eqnarray}
{\bf W}({\bf x},t,t_0)=
\begin{pmatrix} -\beta(t_0)\partial_{t_0} \G^h &  -\G^h \\  
 \beta(t)\beta(t_0)\partial_t\partial_{t_0} \G^h & \beta(t)\partial_t \G^h \end{pmatrix}({\bf x},{\bf 0},t,t_0),\label{eq70HHx}
\end{eqnarray}
with $\G^h({\bf x},{\bf 0},t,t_0)=\G({\bf x},{\bf 0},t,t_0)-\G^a({\bf x},{\bf 0},t,t_0)$.
Given the unified form of equations (\ref{eq1})--(\ref{eq132x}), it follows that equation (\ref{eq70HHx}) also holds for 3D elastodynamic waves in time-variant materials.
Matrix $\G({\bf x},{\bf 0},t,t_0)$ is the $3\times 3$ Green's matrix, i.e., 
the causal response to a unit source  at ${\bf x}={\bf 0}$ and $t_0$, observed at ${\bf x}$ and $t$.
It obeys the causality condition $\G({\bf x},{\bf 0},t,t_0)={\bf O}$ for $t<t_0$.
Furthermore, $\G^a({\bf x},{\bf 0},t,t_0)$ is the acausal Green's matrix, with acausality condition $\G^a({\bf x},{\bf 0},t,t_0)={\bf O}$ for $t>t_0$.
Hence, it is the acausal field observed at ${\bf x}$ and $t$, propagating to a unit sink  
at ${\bf x}={\bf 0}$ and $t_0$, 
after which it vanishes.
The difference of these Green's matrices, $\G^h({\bf x},{\bf 0},t,t_0)$, is called the homogeneous Green's matrix, because the singularities related to the source and the sink at 
${\bf x}={\bf 0}$ and $t_0$ cancel each other.
It should be noted that this straightforward relation  between the propagator matrix and the Green's matrices (equation (\ref{eq70HHx})) is limited to time-variant materials, where the initial condition
for the propagator matrix and the (a)causality conditions for the Green's matrices are all defined in time. In contrast, for space-variant materials, the boundary condition
for the propagator matrix is defined in space, whereas the (a)causality conditions for the Green's matrices are defined in time, so a relation similar to equation (\ref{eq70HHx}) does not exist 
for space-variant materials.

Equations (\ref{eq1}) to (\ref{eq70HHx}) form the basis for the analysis of 3D wave propagation and scattering in time-variant materials and the applications in numerical modeling, 
time-reversed focusing and Green's function retrieval.
To facilitate the analysis, we define the  3D spatial Fourier transformation of a space- and time-dependent quantity ${\bf U}({\bf x},t)$ as
\begin{eqnarray}
\check{\bf U}({\bf k},t) =\int_{{\mathbb{R}}^3} {\bf U}({\bf x},t)\exp (-i{\bf k}\cdot{\bf x}){\rm d}{\bf x},\label{eq12}
\end{eqnarray}
where wave vector ${\bf k}$ is defined as ${\bf k}=(k_x,k_y,k_z)$, $i$ is the imaginary unit, and $\mathbb{R}$ is the set of real numbers.
In the ${\bf k},t$-domain, wave equation (\ref{eq1}) becomes
$\partial_t\check{\bf q}=\check{\bf A}\check{\bf q}+\check{\bf d}$, with $\nabl$ (in matrix ${\bf A}$) replaced by $i{\bf k}$, and
 $\check{\bf A}$ also obeying the symmetry relation of equation (\ref{eqAsymt}). 
The transformed propagator matrix $\check{\bf W}({\bf k},t,t_0)$ obeys $\partial_t\check{\bf W}=\check{\bf A}\check{\bf W}$,
with initial condition $\check{\bf W}({\bf k},t_0,t_0)={\bf I}$. We use these transformed equations to establish some fundamental properties.
 
 First, note that the quantity $\check{\bf q}^\dagger{\bf N}\check{\bf q}$, where  superscript $\dagger$ denotes transposition and complex conjugation, is proportional
 to the net field-momentum density in the direction of ${\bf k}$ \cite{Feynmann63Book2, Burns2020NJP, Wapenaar2026PIER}.
 For the source-free situation we find (using that $\check{\bf A}$ is real-valued)
 \begin{eqnarray}
\partial_t(\check{\bf q}^\dagger{\bf N}\check{\bf q})
=\check{\bf q}^\dagger(\check{\bf A}^t{\bf N}+{\bf N}\check{\bf A})\check{\bf q}=0,\label{eq9}
\end{eqnarray}
which confirms the conservation of net field-momentum density for time-variant materials 
\cite{Morgenthaler58IRE, Mendonca2002PS, Caloz2020IEEE2, Koutserimpas2020IEEE, Wapenaar2026PIER} and extends its validity for 3D elastodynamic waves.

Second, analogous to equation (\ref{eq9}), we find that $\check{\bf W}^t({\bf k},t, t_0){\bf N}\check{\bf W}({\bf k},t, t_N)$ is conserved.
Substituting $t=t_0$ and, subsequently, $t=t_N$, we obtain the following symmetry relation
\begin{eqnarray}
{\bf N}\check{\bf W}({\bf k},t_0,t_N)=\check{\bf W}^t({\bf k},t_N,t_0){\bf N}.\label{eq253}
\end{eqnarray}
The same symmetry relation holds in the ${\bf x},t$-domain. Substituting equation (\ref{eq70HHx}), we thus obtain
\begin{eqnarray}
 \G^h({\bf x},{\bf 0},t_0,t_N)=-\{\G^h({\bf x},{\bf 0},t_N,t_0)\}^t.\label{eq1044K}
\end{eqnarray}
For $t_N>t_0$, using the definition of $\G^h({\bf x},{\bf 0},t,t_0)$ and the (a)causality conditions discussed below equation (\ref{eq70HHx}), this gives
\begin{eqnarray}
\G^a({\bf x},{\bf 0},t_0,t_N)=\G^t({\bf x},{\bf 0},t_N,t_0).\label{eq1044}
\end{eqnarray}
This reciprocity relation states that in a homogeneous, time-variant material, the acausal field, observed at ${\bf x}$ and $t_0$, propagating to a unit sink at ${\bf x}={\bf 0}$ and $t_N$,
is equal to the transposed causal response to a unit source at ${\bf x}={\bf 0}$ and $t_0$, observed at ${\bf x}$ and $t_N$.
This is the counterpart of classical source-receiver reciprocity in an inhomogeneous, time-invariant material. Equation (\ref{eq1044}) will be illustrated with a numerical example later. 

Next, we derive an algorithm for numerical modeling. Using  the spatial Fourier transformation defined in equation (\ref{eq12}), 
the spatial convolution in the ${\bf x},t$-domain in equation (\ref{eq132x}) is replaced by a multiplication in the ${\bf k},t$-domain, 
hence $\check{\bf q}({\bf k},t)=\check{\bf W}({\bf k},t,t_0)\check{\bf q}({\bf k},t_0)$.
By applying this recursively, we obtain the  recursive expression 
\begin{eqnarray}
&&\hspace{-1cm}\check{\bf W}({\bf k},t_N,t_0)\nonumber\\
&&\hspace{-1cm}=\check{\bf W}({\bf k},t_N,t_{N-1})\cdots\check{\bf W}({\bf k},t_n,t_{n-1})\cdots\check{\bf W}({\bf k},t_1,t_0).\label{eq43BB}
\end{eqnarray}
By making small enough time steps, we can take the material parameters time-invariant between $t_{n-1}$ and $t_n$ for all $n$. 
We thus obtain
\begin{eqnarray}
\check{\bf W}({\bf k},t_n,t_{n-1})=\exp\{\check{\bf A}_n\Delta t_n\},\label{eq10}
\end{eqnarray}
with $\Delta t_n=t_n-t_{n-1}$. This expression is discussed in more detail in the Appendix.

\begin{figure}[t]
\centerline{\hspace{4cm}\epsfxsize=11 cm \epsfbox{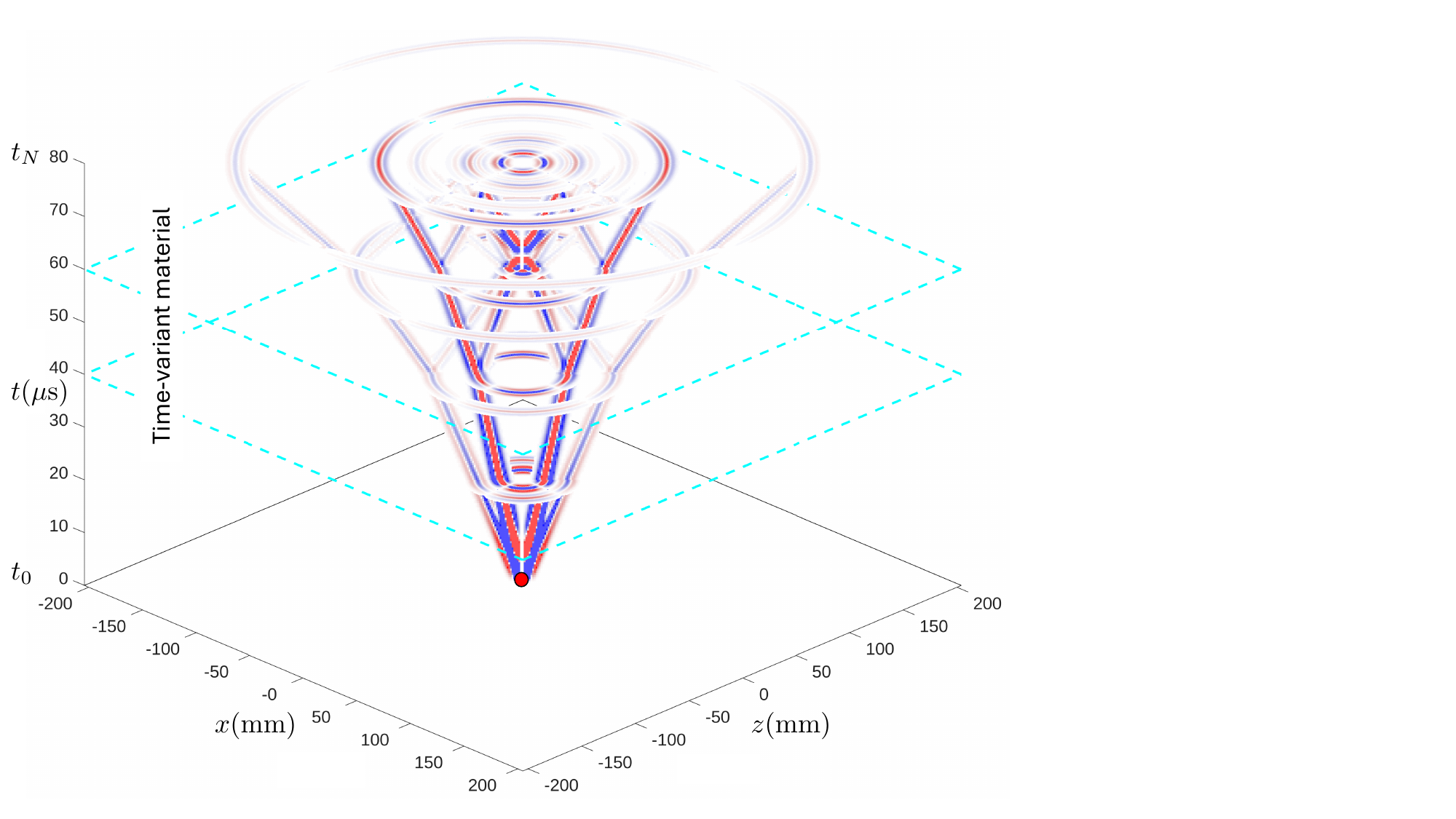}}
\caption{Cross-section at $y=0$ of the 3D causal Green's function ${\cal G}_{x,z}({\bf x},{\bf 0},t,t_0)$ (convolved with a spatial wavelet) in a piecewise constant, time-variant elastic material.
The red dot indicates the source at ${\bf x}={\bf 0}$ and $t_0=0\, \mu$s. The dashed blue planes indicate the time boundaries at $t_n=40\, \mu$s and $t_n=60\, \mu$s.
Movie available at https://www.keeswapenaar.nl/TimeMaterial/Green3D.mp4 
}\label{Figure1}
\end{figure}

\begin{figure}[t]
\centerline{\hspace{16cm}\epsfxsize=24 cm \epsfbox{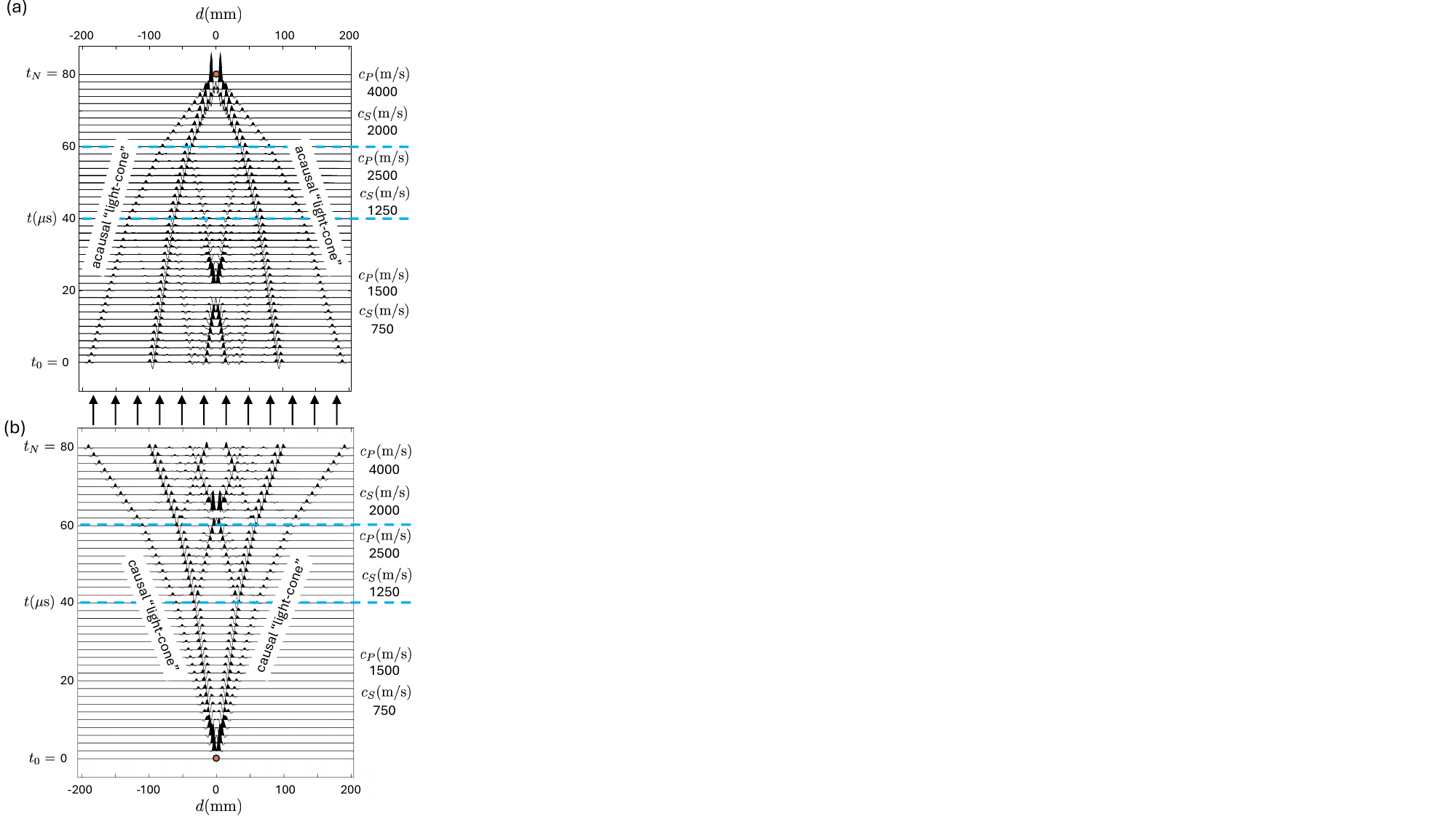}}
\caption{Cross-sections at $y=0$ along the diagonal $x=z$  
of (a) the acausal Green's function ${\cal G}^a_{z,x}({\bf x},{\bf 0},t,t_N)$ and (b) the causal Green's function ${\cal G}_{x,z}({\bf x},{\bf 0},t,t_0)$. }
\label{Figure2}
\end{figure}

As an illustration, we model the propagator matrix in a homogeneous, time-variant elastic material.
The variations of the parameters can, for example, be the result of a time-varying external stress \cite{Verdon2008GEO}. For the illustration, we consider an idealized situation 
with three time-independent slabs, separated by two instantaneous time boundaries.
The duration of the first slab is 40 $\mu$s, and that of the second and third slabs is 20 $\mu$s.
The $P$-wave velocities $c_P$ of the three slabs are 1500, 2500 and 4000 m/s, respectively, the $S$-wave velocities $c_S$ are 750, 1250 and 2000 m/s, respectively, 
and the mass density $\rho$ is constant throughout. Equation (\ref{eq43BB}) is used to model the propagator matrix, with constant time steps $\Delta t= 0.4\,\mu$s.
We start at $t_0=0\,\mu$s and obtain $\check{\bf W}({\bf k},t_n,t_0)$ after $n$ steps. According to equation (\ref{eq70HHx}), the upper-right sub-matrix of $\check{\bf W}({\bf k},t_n,t_0)$ equals
(for $t_n>t_0$) minus the causal Green's matrix $\check\G({\bf k},{\bf 0},t_n,t_0)$. 
Using an inverse 3D spatial Fourier transformation we obtain $\G({\bf x},{\bf 0},t_n,t_0)$. 
We select the component ${\cal G}_{x,z}({\bf x},{\bf 0},t_n,t_0)$, which is the $x$-component of the 
Green's function observed at ${\bf x}$ and $t_n$, in response
to a unit source in the $z$-direction at ${\bf x}={\bf 0}$ and $t_0$. This component of the Green's matrix (convolved with a 3D spatial wavelet
with a central wavenumber $k_0/2\pi=100$ m$^{-1}$) is shown in a 3D display as a function of ${\bf x}=(x,y=0,z)$ and $t$ in Figure \ref{Figure1}.
Figure \ref{Figure2}b shows a cross-section along the diagonal $x=z$ (avoiding the zero-crossings at the $x$- and $z$-axes) of Figure \ref{Figure1} as a function
of $d=\sqrt{x^2+z^2}$ (the distance to the source), for every 5th ``snapshot'' (hence, in this display $\Delta t= 2.0\,\mu$s).
A time- and space-dependent gain is applied in this display to compensate for the rapidly decaying amplitudes due to the 3D spreading.
In the first time slab (from $t_0=0$ to $t_n=40\,\mu$s) we see $P$- and $S$-waves propagating with different velocities, away from the source at ${\bf x}={\bf 0}$. 
These waves get partly transmitted and partly reflected by the  first time boundary  at $t_n=40\, \mu$s. 
Unlike at a space boundary, no conversion between $P$- and $S$-waves occurs at a time boundary. Hence, in the second time slab (from $t_n=40$ to $t_n=60\,\mu$s)
the total number of events (transmitted and reflected) has doubled compared to those in the first time slab.
The second time boundary, at $t_n=60\, \mu$s, causes again a doubling of the number of events. Note that the direct outward propagating $P$-waves form a causal ``light-cone'', originating from the
source at ${\bf x}={\bf 0}$ and $t_0$. All scattering occurs inside this cone.

Next, we model $\check{\bf W}({\bf k},t_n,t_N)$, with constant negative time steps $\Delta t= - 0.4\,\mu$s, starting at $t_N=80\,\mu$s.
According to equation (\ref{eq70HHx}), the upper-right sub-matrix of $\check{\bf W}({\bf k},t_n,t_N)$ equals
(for $t_n<t_N$) the acausal Green's matrix $\check\G^a({\bf k},{\bf 0},t_n,t_N)$. Figure \ref{Figure2}a shows, for fixed $t_N$, a cross-section at $y=0$ along the diagonal $x=z$ of every 5th snapshot
of the component ${\cal G}^a_{z,x}({\bf x},{\bf 0},t_n,t_N)$  (convolved with a spatial wavelet). From bottom to top, i.e., for increasing $t$, we observe a complex acausal wave field,
converging to the sink at ${\bf x}={\bf 0}$ and $t_N$.
Partial transmission and reflection occurs at each time boundary, after which the total number of events is halved.
The direct inward propagating $P$-waves form an acausal ``light-cone'' and all scattering occurs inside this cone.
Despite the fact that Figure \ref{Figure2}a is not the time-reversal of Figure \ref{Figure2}b, 
we observe that the snapshot at $t_0$ in Figure \ref{Figure2}a, ${\cal G}^a_{z,x}({\bf x},{\bf 0},t_0,t_N)$, is identical to the  
snapshot at $t_N$ in Figure \ref{Figure2}b, ${\cal G}_{x,z}({\bf x},{\bf 0},t_N,t_0)$. This confirms the reciprocity relation of equation (\ref{eq1044}).

Figure \ref{Figure2} as a whole resembles the principle of classical time-reversed focusing:
it suggests that when the response at $t_N$ (Figure \ref{Figure2}b), is emitted by sources at $t_0$ 
into a repetition of the original (non-reversed) time-variant material, it focuses at ${\bf x}={\bf 0}$ and $t_N$ (Figure \ref{Figure2}a).
However, to make this happen, the response at $t_N$ needs to be manipulated in a specific way before it is re-emitted  into the material.
To see this, consider the following special case of equation (\ref{eq43BB})
\begin{eqnarray}
\check{\bf W}({\bf k},t_n,t_N)=\check{\bf W}({\bf k},t_n,t_0)\check{\bf W}({\bf k},t_0,t_N).\label{eq79KK}
\end{eqnarray}
Using the symmetry relation of equation (\ref{eq253}) and transforming the resulting expression to the ${\bf x},t$-domain, we obtain
\begin{eqnarray}
{\bf W}({\bf x},t_n,t_N)={\bf W}({\bf x},t_n,t_0)*{\bf N}^{-1}{\bf W}^t({\bf x},t_N,t_0){\bf N}.\label{eq100GG}
\end{eqnarray}
We choose $t_0<t_n< t_N$. Substituting equation (\ref{eq70HHx}),
using the definition of $\G^h({\bf x},{\bf 0},t,t_0)$ and the (a)causality conditions discussed below equation (\ref{eq70HHx}), this gives
for the upper-right sub-matrix of ${\bf W}({\bf x},t_n,t_N)$
\begin{eqnarray}
&&\hspace{-1cm}{\G}^a({\bf x},{\bf 0},t_n,t_N)=\nonumber\\
&&\hspace{-0.5cm}-\mu(t_0)\Bigl(\{\partial_{t_0}{\G}({\bf x},{\bf 0},t_n,t_0)\}*
{\G}^t({\bf x},{\bf 0},t_N,t_0)\nonumber\\
&&\hspace{0.2cm}-{\G}({\bf x},{\bf 0},t_n,t_0)*\partial_{t_0}{\G}^t({\bf x},{\bf 0},t_N,t_0)\Bigr).\label{eq91GGx}
\end{eqnarray}
The right-hand side states that the transposed of the response ${\G}({\bf x},{\bf 0},t_N,t_0)$ (Figure \ref{Figure2}b) and its derivative are
 emitted into the same material by sources at $t_0$ and propagated by ${\G}({\bf x},{\bf 0},t_n,t_0)$ and its derivative from $t_0$ to $t_n$. This results in the 
acausal Green's function ${\G}^a({\bf x},{\bf 0},t_n,t_N)$ on the left-hand side (Figure \ref{Figure2}a), which focuses  at ${\bf x}={\bf 0}$ and $t_N$. 
Hence, equation (\ref{eq91GGx}) represents the counterpart of classical time-reversed focusing. 

For inhomogeneous, time-invariant materials, the principle of time-reversed focusing is closely linked to the principle of Green's function retrieval by temporal cross-correlation
\cite{Derode2003JASA, Wapenaar2005JASA}. Here we show a similar link for homogenous, time-variant materials. To this end,
we consider again equation (\ref{eq100GG}), but this time we drop the condition $t_n<t_N$. This implies that the left-hand side of equation (\ref{eq91GGx}) is replaced by minus the
homogeneous Green's function ${\G}^h({\bf x},{\bf 0},t_n,t_N)$. On the right-hand side of equation (\ref{eq91GGx}) we replace ${\G}^t({\bf x},{\bf 0},t_N,t_0)$ and its derivative
by ${\G}^t(-{\bf x},{\bf 0},t_N,t_0)$ and its derivative, which is allowed because the Green's functions are symmetric in ${\bf x}$.
Now the terms on the right-hand side can be interpreted as spatial cross-correlations of wave fields and their derivatives, observed by receivers at $t_n$ and $t_N$, in response to sources at $t_0$.
This gives minus the homogeneous Green's function ${\G}^h({\bf x},{\bf 0},t_n,t_N)$ on the left-hand side, which, for $t_n>t_N$, is the response to a source at $t_N$, observed at $t_n$.
In other words, by spatial cross-correlation, one of the receivers (at $t_N$) is turned into a virtual source, of which the response is observed by the other receiver (at $t_n$).
Hence, this modification of equation (\ref{eq91GGx}) represents the counterpart of classical Green's function retrieval
 \cite{Weaver2001PRL, Campillo2003Science, Wapenaar2004PRL, Snieder2004PRE, VanManen2005PRL}.


In conclusion, the unified formalism presented here not only forms a solid theoretical basis for establishing several fundamental aspects of
3D wave propagation and scattering in time-variant materials, 
it also forms the basis for the application of 3D electromagnetic and elastodynamic time-reversed focusing and Green's function retrieval in time-variant materials.
The analysis in this study is restricted to time-variant materials, of which the parameters do not vary with position. The extension to 3D arbitrarily inhomogeneous, time-variant materials
forms an exciting subject for further research.

{\it Data availability}---No data were analyzed in this study.

{\it Acknowledgments}---D.-J.v.M. was supported by Swiss National Science Foundation Grant No. 197182.


%

\appendix
\newpage

\section{}\label{AppA}

We start with deriving equations (\ref{eq1}) and (\ref{eq2}) in the main text for electromagnetic waves in a homogeneous, isotropic, time-variant material.
The Maxwell equations for electromagnetic waves read \cite{Landau60Book, Hoop95Book}
\begin{eqnarray}
\partial_t{\bf D}&=&\nabl\times{\bf H}-{\bf J}^{\rm e},\label{eqA1}\\
\partial_t{\bf B}&=&-\nabl\times{\bf E}-{\bf J}^{\rm m},\label{eqA2}
\end{eqnarray}
where ${\bf E}({\bf x},t)$ and ${\bf H}({\bf x},t)$ are the electric and magnetic field strengths, respectively, as a function of position (${\bf x}$) and time ($t$). 
The other quantities are specified in Table \ref{table1}. The field quantities in these equations are related via the constitutive equations
\begin{eqnarray}
{\bf D}&=&\varepsilon{\bf E},\label{eqA3}\\
{\bf B}&=&\mu{\bf H}.\label{eqA4}
\end{eqnarray}
For this study, we assume that the permittivity and permeability are time-dependent functions:  $\varepsilon(t)$ and $\mu(t)$.
We use equations (\ref{eqA3}) and (\ref{eqA4}) to eliminate  ${\bf E}$ and ${\bf H}$ from equations (\ref{eqA1}) and (\ref{eqA2}) 
\cite{Budko2009PRA, Hoop2014WM, Koutserimpas2020IEEE, Stumpf2026SR, Wapenaar2026PIER}. This yields
\begin{eqnarray}
\partial_t{\bf D}&=&\frac{1}{\mu(t)}\nabl\times{\bf B}-{\bf J}^{\rm e},\label{eqA5}\\
\partial_t{\bf B}&=&-\frac{1}{\varepsilon(t)}\nabl\times{\bf D}-{\bf J}^{\rm m}.\label{eqA6}
\end{eqnarray}
Equations (\ref{eqA5}) and (\ref{eqA6}) can be captured in the form of equation (\ref{eq1}), with ${\bf q}$ and ${\bf d}$ being $6\times 1$ vectors and ${\bf A}$ a $6\times 6$ 
matrix. This forms a useful starting point for modeling the time evolution of ${\bf D}$ and ${\bf B}$ through a homogeneous, time-variant material \cite{Stumpf2026SR}.
Unfortunately, in this case matrix ${\bf A}$ does not obey the symmetry relation, formulated by equation (\ref{eqAsymt}).
To remedy this, we apply minus the curl operator to both sides of equation (\ref{eqA6}), which yields 
\begin{eqnarray}
\partial_t(-\nabl\times{\bf B})=\frac{1}{\varepsilon(t)}\nabl\times\nabl\times{\bf D}+
\nabl\times{\bf J}^{\rm m}.\label{eqA7}
\end{eqnarray}
Equations (\ref{eqA5}) and (\ref{eqA7}) can be captured in the form of equations (\ref{eq1}) and (\ref{eq2}), with the quantities in ${\bf q}$, ${\bf d}$ and ${\bf A}$ specified in Table \ref{table1},
with ${\bf A}$ obeying the symmetry property, formulated by equation (\ref{eqAsymt}).

Next, we derive equations (\ref{eq1}) and (\ref{eq2}) for elastodynamic waves in a homogeneous, isotropic, time-variant material.
The equation for equilibrium of mechanical momentum and the deformation equation \cite{Hoop95Book} read
\begin{eqnarray}
\partial_t{\bf m}&=&\nabl\cdot\taub+{\bf f},\label{eqB1}\\
\partial_t{\bf e}&=&\frac{1}{2}\bigl(\nabl{\bf v}+(\nabl{\bf v})^t\bigr)-{\bf h},\label{eqB2}
\end{eqnarray}
where ${\bf v}({\bf x},t)$ and $\taub({\bf x},t)$ are the particle velocity and the stress tensor, respectively. The other quantities are specified in Table \ref{table1}.
Note that $\taub=\taub^t$, ${\bf e}={\bf e}^t$ and ${\bf h}={\bf h}^t$. The field quantities in equations (\ref{eqB1}) and (\ref{eqB2}) are related via the constitutive equations
\begin{eqnarray}
{\bf m}&=&\rho{\bf v},\label{eqB3}\\
\taub&=&2\mu{\bf e}+\lambda\Theta{\bf I},\label{eqB4}
\end{eqnarray}
with the cubic dilatation $\Theta$ defined as $\Theta={\rm tr}({\bf e})$.
For this study, we assume that the mass density and the Lam\'e parameters are time-dependent functions:  $\rho(t)$, $\lambda(t)$ and $\mu(t)$.
We use equations (\ref{eqB3}) and (\ref{eqB4}) to eliminate  ${\bf v}$ and $\taub$ from equations (\ref{eqB1}) and (\ref{eqB2}). This yields
\begin{eqnarray}
\partial_t{\bf m}&=&2\mu\nabl\cdot{\bf e}+\lambda\nabl\Theta+{\bf f},\label{eqB5}\\
\partial_t{\bf e}&=&\frac{1}{2\rho}\bigl(\nabl{\bf m}+(\nabl{\bf m})^t\bigr)-{\bf h}.\label{eqB6}
\end{eqnarray}
Applying the divergence operator to both sides of equation (\ref{eqB6}) yields
\begin{eqnarray}
\partial_t\nabl\cdot{\bf e}=\frac{1}{2\rho}\nabl\cdot\bigl(\nabl{\bf m}+(\nabl{\bf m})^t\bigr)
-\nabl\cdot{\bf h}.
\end{eqnarray}
Using
\begin{eqnarray}
\nabl\cdot(\nabl{\bf m})&=&
(\nabl\cdot\nabl){\bf m}\nonumber\\
&=&
\nabl(\nabl\cdot{\bf m})-\nabl\times\nabl\times{\bf m},\\
\nabl\cdot(\nabl{\bf m})^t&=&\nabl(\nabl\cdot{\bf m}),
\end{eqnarray}
we obtain
\begin{eqnarray}
\hspace{-.5cm}\partial_t\nabl\cdot{\bf e}=
\frac{1}{2\rho}\bigl(2\nabl(\nabl\cdot{\bf m})-\nabl\times\nabl\times{\bf m}\bigr)
-\nabl\cdot{\bf h}.\label{eqB17}
\end{eqnarray}
Equations (\ref{eqB5}) and (\ref{eqB17}) form a system of two equations for the three wave field vectors ${\bf m}$, $\nabl\cdot{\bf e}$ and $\nabl\Theta$, hence, we need a third equation.
Taking the trace of all terms of  equation (\ref{eqB6}), we obtain
\begin{eqnarray}
\partial_t\Theta=\frac{1}{\rho}\nabl\cdot{\bf m}-q,\label{eqB7}
\end{eqnarray}
with the volume-injection rate density $q$ defined as $q={\rm tr}({\bf h})$. Applying the gradient operator to both sides of equation (\ref{eqB7}) yields
\begin{eqnarray}
\partial_t\nabl\Theta=\frac{1}{\rho}\nabl(\nabl\cdot{\bf m})-\nabl q.\label{eqB12}
\end{eqnarray}
Equations (\ref{eqB5}), (\ref{eqB17}) and (\ref{eqB12})  can be captured in the form of equation (\ref{eq1}), with 
${\bf q}$ and ${\bf d}$ being $9\times 1$ vectors and ${\bf A}$ a $9\times 9$ 
matrix. This forms a useful starting point for modeling the time evolution of ${\bf m}$, $\nabl\cdot{\bf e}$ and $\nabl\Theta$ through a homogeneous, time-variant material.
Unfortunately, in this case matrix ${\bf A}$ does not obey the symmetry relation, formulated by equation (\ref{eqAsymt}).
To remedy this, from here onward we assume that the ratio $\frac{\lambda(t)}{\mu(t)}$ is time-invariant. 
This allows us to combine equations (\ref{eqB17}) and (\ref{eqB12}) into 
\begin{eqnarray}
\hspace{-.6cm}\partial_t\Bigl(2\nabl\cdot{\bf e}+\frac{\lambda}{\mu}\nabl\Theta\Bigr)=\frac{1}{\rho}\D{\bf m}-2\nabl\cdot{\bf h}-\frac{\lambda}{\mu}\nabl q,\label{eqB22}
\end{eqnarray}
where
\begin{eqnarray}
\D&=&\frac{c_P^2}{c_S^2}\nabl\nabl^t-\nabl\times\nabl\times,
\end{eqnarray}
with $P$- and $S$-wave velocities $c_P(t)$ and $c_S(t)$ defined as
\begin{eqnarray}
c_P=\sqrt{\frac{\lambda+2\mu}{\rho}}\quad\mbox{and}\quad c_S=\sqrt{\frac{\mu}{\rho}}.\label{eqC23}
\end{eqnarray}
The assumption that $\frac{\lambda(t)}{\mu(t)}$ is time-invariant implies that $\frac{c_P(t)}{c_S(t)}$ is assumed time-invariant.
Equations (\ref{eqB5}) and (\ref{eqB22}) can be captured in the form of equations (\ref{eq1}) and (\ref{eq2}), with the quantities in ${\bf q}$, ${\bf d}$ and ${\bf A}$ specified in Table \ref{table1},
with ${\bf A}$ obeying the symmetry property, formulated by equation (\ref{eqAsymt}).

Finally, for the elastodynamic situation, we derive  an explicit expression for $\check{\bf W}({\bf k},t_n,t_{n-1})$ for a time-invariant slab between $t_{n-1}$ and $t_n$, hence,
$\check{\bf W}({\bf k},t_n,t_{n-1})=\exp\{\check{\bf A}_n\Delta t_n\}$, with $\Delta t_n=t_n-t_{n-1}$  (equation \ref{eq10}).
 For convenience, from here onward we drop the subscripts $n$ on the right-hand side of this expression. Matrix $\check{\bf A}$ is the Fourier transform of
matrix ${\bf A}$, defined in equation (\ref{eq2}), with the transformed operator $\check\D$ defined as $\check\D=-\frac{c_P^2}{c_S^2}{\bf k}{\bf k}^t+{\bf k}\times{\bf k}\times$, with
\begin{eqnarray}
\hspace{-1.3cm}&&{\bf k}{\bf k}^t=\begin{pmatrix}
k_x^2&k_xk_y&k_xk_z\\
k_xk_y&k_y^2&k_yk_z\\
k_xk_z&k_yk_z&k_z^2\\
\end{pmatrix},\\
\hspace{-1.3cm}&&{\bf k}\times{\bf k}\times=\begin{pmatrix}
-(k_y^2+k_z^2) & k_xk_y & k_xk_z\\
k_xk_y & -(k_x^2+k_z^2) & k_yk_z\\
k_xk_z & k_yk_z & -(k_x^2+k_y^2)
\end{pmatrix}.
\end{eqnarray}
The eigenvalue decomposition of matrix $\check{\bf A}$ reads $\check{\bf A}=\check{\bf L}\check{\bf \Lambda}\check{\bf L}^{-1}$.
Due to the anti block-diagonal structure of matrix $\check{\bf A}$, we can write
\begin{eqnarray}
\hspace{-.4cm}\check{\bf A}=
\underbrace{\begin{pmatrix}\check{\bf L}_U&\check{\bf L}_U\\\check{\bf L}_V&-\check{\bf L}_V\end{pmatrix}}_{\check{\bf L}}
\underbrace{\begin{pmatrix}\check{\bf \Lambda}_1&{\bf O}\\{\bf O}&-\check{\bf \Lambda}_1\end{pmatrix}}_{\check{\bf \Lambda}}
\underbrace{\frac{1}{2}\begin{pmatrix}\check{\bf L}_U^{-1}&\check{\bf L}_V^{-1}\\\check{\bf L}_U^{-1}&-\check{\bf L}_V^{-1}\end{pmatrix}}_{{\check{\bf L}}^{-1}},\label{eqC26}
\end{eqnarray}
or
%
%
\begin{eqnarray}
\begin{pmatrix}{\bf O} & -\frac{1}{\beta(t)}{\bf I} \\
-\frac{1}{\alpha(t)}\check\D& {\bf O}\end{pmatrix}=
\begin{pmatrix}{\bf O} & \check{\bf L}_U\check{\bf \Lambda}_1\check{\bf L}_V^{-1} \\
\check{\bf L}_V\check{\bf \Lambda}_1\check{\bf L}_U^{-1}& {\bf O}\end{pmatrix}.\label{eq63}
\end{eqnarray}
This equation is solved by
\begin{eqnarray}
&&\hspace{-1.2cm}\check{\bf L}_U=\begin{pmatrix}
k_x & -k_y	 & -k_z \\
k_y &  k_x & 0     \\
k_z & 0      &  k_x  
\end{pmatrix},
\check{\bf \Lambda}_1=-ik_r\begin{pmatrix}
c_P & 0 & 0 \\
0 & c_S & 0 \\
0 & 0 & c_S 
\end{pmatrix},\\
&&\hspace{-1.2cm}\check{\bf L}_V=ik_r\beta\begin{pmatrix}
c_Pk_x &-c_Sk_y & -c_Sk_z\\
c_Pk_y &c_Sk_x & 0\\
c_Pk_z &0 &c_Sk_x
\end{pmatrix},\label{eqC14}
\\\nonumber
\end{eqnarray}
with $k_r=|{\bf k}|=\sqrt{k_x^2+k_y^2+k_z^2}$.
Using the eigenvalue decomposition of $\check{\bf A}$, we express the propagator matrix as
\begin{eqnarray}
\check{\bf W}({\bf k},t_n,t_{n-1})=\exp\{\check{\bf A}\Delta t\}=\check{\bf L}\exp\{\check{\bf \Lambda}\Delta t\}\check{\bf L}^{-1}.\label{eqC25}
\end{eqnarray}
We partition $\check{\bf W}({\bf k},t_n,t_{n-1})$ as
\begin{eqnarray}
\check{\bf W}({\bf k},t_n,t_{n-1})=
\begin{pmatrix} \check{\bf W}^{U,U} &  \check{\bf W}^{U,V} \\   \check{\bf W}^{V,U} &  \check{\bf W}^{V,V} \end{pmatrix}({\bf k},t_n,t_{n-1}).\label{eq31}
\end{eqnarray}
Substituting the definitions of $\check{\bf L}$, $\check{\bf\Lambda}$ and $\check{\bf L}^{-1}$ of equation (\ref{eqC26}) into equation (\ref{eqC25}), we obtain
\begin{eqnarray}
\hspace{-.8cm}\check{\bf W}^{U,U}({\bf k},t_n,t_{n-1}) &=&\check{\bf L}_U\cos\bigl(i\check{\bf\Lambda}_1\Delta t\bigr)\check{\bf L}_U^{-1},\label{eqC32}\\
\hspace{-.8cm}\check{\bf W}^{U,V}({\bf k},t_n,t_{n-1}) &=&-i\check{\bf L}_U\sin\bigl(i\check{\bf\Lambda}_1\Delta t\bigr)\check{\bf L}_V^{-1},\label{eqC33}\\
\hspace{-.8cm}\check{\bf W}^{V,U}({\bf k},t_n,t_{n-1}) &=&-i\check{\bf L}_V\sin\bigl(i\check{\bf\Lambda}_1\Delta t\bigr)\check{\bf L}_U^{-1},\\
\hspace{-.8cm}\check{\bf W}^{V,V}({\bf k},t_n,t_{n-1}) &=&\check{\bf L}_V\cos\bigl(i\check{\bf\Lambda}_1\Delta t\bigr)\check{\bf L}_V^{-1},\label{eqC35}
\end{eqnarray}
where
\begin{eqnarray}
\cos\bigl(i\check{\bf\Lambda}_1\Delta t\bigr)&=&\begin{pmatrix}
\cos\phi & 0 & 0\\
0 & \cos\psi & 0\\
0 & 0 & \cos\psi \end{pmatrix},\\
\sin\bigl(i\check{\bf\Lambda}_1\Delta t\bigr)&=&\begin{pmatrix}
\sin\phi & 0 & 0\\
0 & \sin\psi & 0\\
0 & 0 & \sin\psi \end{pmatrix},
\end{eqnarray}
with
\begin{eqnarray}
\phi=k_rc_P\Delta t,\quad \psi=k_rc_S\Delta t.
\end{eqnarray}
The limits for $k_r\to 0$ of equations (\ref{eqC32}) -- (\ref{eqC35}) are finite. Expressions for the electromagnetic situation are obtained by taking the limit 
$c_P\to 0$ of the above expressions and replacing $c_S$ by the electromagnetic propagation velocity $c_{EM}$, defined as
\begin{eqnarray}
c_{EM}=\frac{1}{\sqrt{\varepsilon\mu}}\label{eqC42}
\end{eqnarray}
(where  $\mu$ in equation (\ref{eqC42}) is not the same as $\mu$ in equation (\ref{eqC23}), see Table \ref{table1}).
Note that, although $\check{\bf L}_V$ is not invertible for $c_P=0$, the limits for $c_P\to 0$ of equations (\ref{eqC33}) and (\ref{eqC35}) are finite.

\end{document}